\documentclass[graybox,natbib,smallextended]{svmult}

\usepackage{amsmath}%
\usepackage{amssymb}%
\usepackage{wasysym}%

\usepackage{mathptmx}       
\usepackage{helvet}             
\usepackage{courier}           
\usepackage{type1cm}         
\usepackage{makeidx}              
\usepackage{graphicx}              
\usepackage{multicol}                
\usepackage[bottom]{footmisc}  

\makeindex             

\usepackage{color}
\usepackage{hyperref}
\usepackage{url}
\usepackage{soul} 
\usepackage{ulem} 

\def\pc#1{principal component#1 (PC#1)\gdef\pc{PC}}
\def\nr#1{numerical relativity#1 (NR#1)\gdef\nr{NR}}
\def\gw#1{gravitational wave#1 (GW#1)\gdef\gw{GW}}

\def\imbh#1{intermediate mass black hole#1(IMBH#1)\gdef\imbh{IMBH}}
\def\smbh#1{supermassive black hole#1(SMBH#1)\gdef\smbh{SMBH}}
\def\bbh#1{binary black hole#1 (BBH#1)\gdef\bbh{BBH}}
\def\bh#1{black hole#1 (BH#1)\gdef\bh{BH}}
\def\ns#1{neutron star#1 (NS#1)\gdef\ns{NS}}
\def\gw#1{gravitational wave#1 (GW#1)\gdef\gw{GW}}
\def\pnw#1{post-Newtonian#1 (PN#1)\gdef\pnw{PN}}
\def\eos#1{equation of state#1 (EOS#1)\gdef\eos{EOS}}
\def\grb#1{gamma-ray burst#1 (GRB#1)\gdef\grb{GRB}}
\def\amr#1{adaptive mesh refinement#1 (AMR#1)\gdef\amr{AMR}}
\def\isco#1{innermost stable circular orbit#1 (ISCO#1)\gdef\isco{ISCO}}

\begin{document}

\title*{Investigating Binary Black Hole Mergers with Principal Component Analysis}
\author{J. Clark\inst{1}, L. Cadonati\inst{1,2}, J. Healy\inst{4},  I.S. Heng\inst{3}, J. Logue\inst{3}, N. Mangini\inst{1}, L. London\inst{4}, L. Pekowsky \inst{4}, D. Shoemaker\inst{4}}
\authorrunning{J. Clark et al}
\institute{
\inst{1} University of Massachusetts Amherst, Amherst, MA 01003, USA \\ 
\inst{2} Cardiff University, Cardiff, CF24 3AA, United Kingdom \\
\inst{3} SUPA, School of Physics and Astronomy, University of Glasgow, G12 8QQ, United Kingdom \\
\inst{4} Center for Relativistic Astrophysics, Georgia Institute of Technology, Atlanta GA 30332}
            

\maketitle

\abstract{Despite recent progress in numerical simulations of the coalescence of binary black hole systems,
 highly asymmetric spinning systems and the construction of accurate  physical templates remain challenging and computationally expensive.
We explore the feasibility of a prompt and robust test of whether the signals exhibit evidence for generic features that can educate new simulations.
%
We form catalogs of numerical relativity waveforms with distinct physical effects and compute the relative probability that a gravitational wave signal belongs to each catalog. We introduce an algorithm designed to perform this task for coalescence signals 
 using principal component analysis of waveform catalogs and Bayesian model selection and demonstrate its effectiveness.}

\section{Introduction}
\label{intro}

The coalescence of two black holes is arguably the most powerful source of \gw{s} detectable by the second generation of ground based detectors: Advanced LIGO~\citep{Harry:2010zz}, Advanced Virgo~\citep{Acernese:2009}, and KAGRA~\citep{Somiya:2011np}. 
The discovery of these signatures, forecast within the next few years~\citep{Aasi:2013wya}, will open a new era of gravitational wave astrophysics, where the \gw{} signature will provide insights on the physics 
of the source.   

To decode the information in the \gw{} waveform, we need a careful mapping with the masses and the spin magnitude and orientation of the black holes; this is the charge of \nr{}. 
 While  available \nr{} waveforms span an increasing portion of the physical
 parameter space of unequal mass, spin and precessing  \bbh{s}~\citep{Ajith:2012az,Hinder:2013oqa},  each simulation takes a week or more to run. A complete coverage of the full parameter space remains a slow but important endeavor to enable \gw{} matched filtering and parameter estimation~\citep{Thorne:87,Aasi:2013jjl}. 
 
 The LIGO and Virgo Collaborations have refined techniques for the search of generic \gw{} transients, or {\em bursts}, which don't assume a specific waveform but rely on a coherent \gw{}  in multiple detectors for a variety of plausible sources~\citep{Abadie:2012rq, 
  Andersson:2013mrx}. The work presented here aims to answer the question of how a transient detected by a template-less burst search can trigger new \nr{}  simulations in interesting regions of the \bbh{} parameter space. 
 We introduce a proof-of-concept study, which uses the method of Principal Component Analysis (PCA) to compare a plausible signal to catalogs of \nr{} waveforms, which represent certain regions of the  \bbh{}  physical parameter space.

\section{Binary Black Hole Merger Simulations}
\label{bbh}

\begin{table} [b]
\caption{Physical parameters for the three catalogs used in this study.}
\label{t:catalog}       
\begin{tabular}{p{3.5cm}p{2cm}p{2cm}p{2cm}}
\hline\noalign{\smallskip}
Name & Q & HR & RO3   \\
\noalign{\smallskip}\svhline\noalign{\smallskip}
Mass Ratio, $q=m_1/m_2$ & 1 -- 2.5 & 1 -- 4 & 1.5 -- 4 \\
Spin magnitude, $a$ & 0.0 & $0.0-0.9$ & 0.4, 0.6 \\
Tilt Angle, $\theta$ & 0.0 & 0.0 & $45^o-270^o$\\
N waveforms & 13  & 15 & 20 \\
\noalign{\smallskip}\hline\noalign{\smallskip}
\end{tabular}
\end{table}

The \gw{} waveform produced by solar and intermediate mass \bbh{} systems spans the sensitive band of ground based detectors through the inspiral, merger and ringdown phases. While post-Newtonian and perturbation theories adequately describe the inspiral and ringdown, numerical relativity is necessary to capture the physics of the merger.  \nr{} has been probing 
the parameter space of binary black hole mergers since the breakthrough of 2005~\citep{Pretorius:2005gq} achieving extreme mass ratios~\citep{Lousto:2010ut}, extreme spin magnitudes~\citep{Lovelace:2011nu} and
many precessing runs~\citep{Mroue:2013xna,Pekowsky:2013ska}.  

The \nr{} waveforms used in this paper were produced by the \textsc{Maya} code of the Georgia Institute of Technology~\citep{Vaishnav:2007nm}
The \textsc{Maya} code uses the \texttt{Einstein Toolkit}\footnote{\url{http://www.einsteintoolkit.org}},
which is based on the \texttt{CACTUS}\footnote{\url{http://www.cactus code.org}} infrastructure
and \texttt{CARPET}  mesh refinement \citep{Schnetter-etal-03b}.  The output of all simulations is the Weyl Scalar, $\Psi_4$,
decomposed into spin-weighted spherical harmonics that is then converted to strain~\citep{Reisswig:2010di}.  

For this work we use 48 NR runs, listed in Table~\ref{t:catalog} without hybridization with post-Newtonian waveforms.  The Q-series contains 13 non-spinning, unequal-mass simulations. 
We use 15 runs from the HR-series, a set of unequal-mass, equal spin simulations, with initial spin parallel to the initial angular momentum.
The RO3-series is a set of 20 unequal-mass simulations with the lighter black hole spin aligned to the initial angular momentum (z-axis) and the other black hole at a tilt angle $\theta$ with the z-axis in the xz-plane;
%
%
%
these systems are precessing and the tilt-angles are defined at a specific separation of the black holes at one instant in the evolution of the binary system and change in time. 
While the runs are tabulated with initial parameters,  there is no functional form to relate one waveform to the next;
we use a Principal Component Analysis to determine the main features of  each catalog. 

\section{Principal Component Analysis and Bayesian Model Selection}
\label{pca}

\begin{figure}[h]
\sidecaption
\includegraphics[scale=.9]{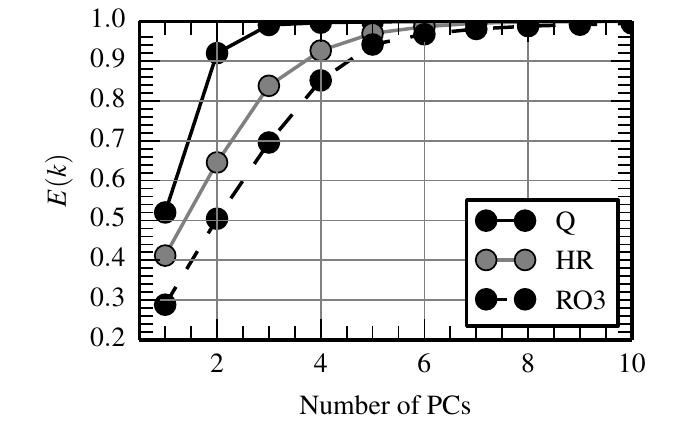}
\caption{Cumulative eigenvector energy as a function of the number of principal components for the three catalogs  in this study. We use the number of PCs that provides 90\% of the energy:  2 PCs for set Q, 4 PCs for set HR and 5 PCs for set RO3. }
\label{f:npca}       
\end{figure}

\begin{figure}[h]
\includegraphics{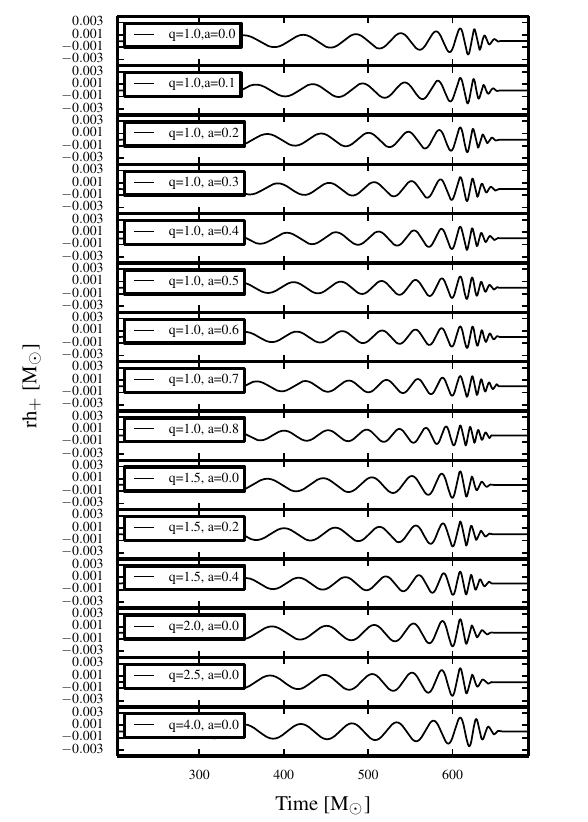}
\includegraphics{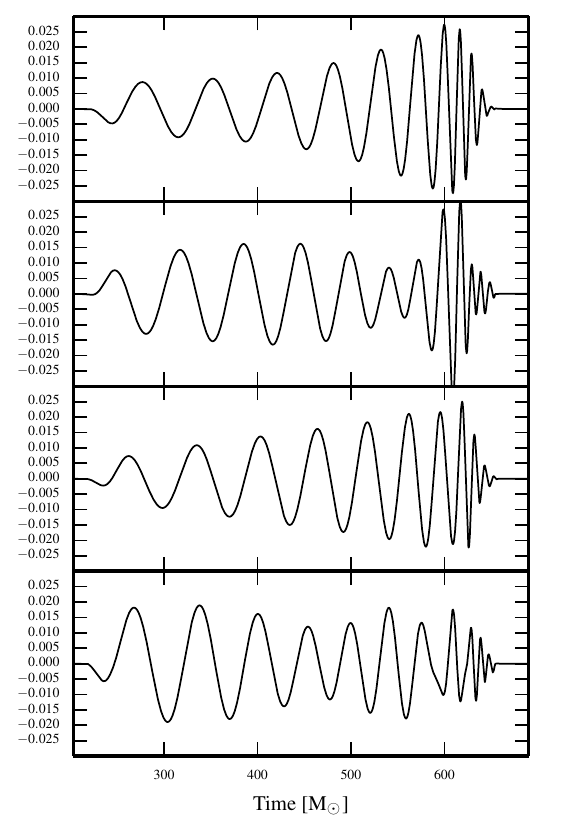}
\caption{\emph{Left}: the waveforms in the HR catalog of spinning, non
precessing waveforms used in this study. \emph{Right}: The principle component
decomposition of the HR catalog.}
\label{f:HRcatalog}       
\end{figure}

We parametrize the \nr{} waveform catalogs of \S\ref{bbh} with an orthonormal set of \pc{s},  
obtained with a standard singular value decomposition~\citep{heng:09,roever:09}. 
For a catalog of $n$ waveforms $\{h_i\}_{i =1,..,n}$ with $m$ samples, we create  a matrix {\bf H} whose columns corresponds to each waveform. 
We then factorize the resulting $m \times n$  matrix {\bf H} so that:
\begin{equation}
\label{eq:svd}
{\bf H} = {\bf USV}^T,
\end{equation}
where {\bf U} is an $m \times m$ matrix whose columns are the eigenvectors of ${\bf HH}^T$ and {\bf V} is an $n \times n$ matrix whose columns are eigenvectors of ${\bf H}^T{\bf H}$.
 The  $m \times n$ matrix  ${\bf S}$ will have all zeros, except for the 
 $\{S_{jj}\}_{j =1,..,n}$ terms, which correspond to the square root of the $j$th eigenvalue. 
%
%
%
 ${\bf U}$ contains the catalog's  \pc{s}, 
ranked by their corresponding eigenvalue:
the first column is the first \pc{}, which encapsulates
 the most significant  features common to all waveforms in the
catalog, the second column, corresponding to  the second largest eigenvalue,
 describes the second most significant common features in the catalog, and so on. 
The waveforms in {\bf H} can be reconstructed as a linear combination of \pc{s}:
\begin{equation}
\label{eq:sig}
h_i \approx \sum_{j=1}^{k}{U_j}{\beta_j}\,\,,
\end{equation}
where $h_i$ is the catalog waveform, $u_j$ is the
$j$th \pc{}  and $\beta_j$ is the corresponding 
coefficient,  obtained by projecting $h_i$ onto $u_j$. The
sum over $k$ \pc{s} is an approximation of the desired waveform, since in general
$k<n$.
In this analysis, the choice of $k$ is determined by the {\it cumulative eigenvalue energy}, $E(k)$, 
shown in Figure~\ref{f:npca}:
\begin{equation}
\label{eg:eigenvalues}
E(k) = \frac{\sum_{i=1}^k S_{ii}}{\sum_{j=1}^n S_{jj}}
\end{equation}
%
In this analysis we use $k$ PCs, so that
$E(k) \ge 0.9$. This corresponds to 2, 4 and 5 PCs for the Q, HR and RO3 catalogs respectively.
A selection of the waveforms from the HR catalog and corresponding PCs are shown in Figure~\ref{f:HRcatalog}. 

%

%
%
%
Following the seminal work on Burst signals in~\citep{clark:07,Logue:2012zw}, 
the \pc{s}  can be used to identify generic features for a
measured waveform through the posterior odds ratio, which is widely used
in \gw{} data analysis to compare the probabilities of two competing models
$M_i$ and $M_j$.  Given data $D$,  the odds ratio
${\mathcal O}_{ij}$ is  the ratio of posterior probabilities for each
model:
\begin{equation}
{\mathcal O}_{ij}=\frac{p(M_i)}{p(M_j)}\frac{p(D|M_i)}{p(D|M_j)} = \pi_{ij}\frac{Z_i}{Z_j}\,\,,
\end{equation}
where $\pi_{ij}$ is the \emph{prior odds ratio} which reflects any bias one has
for the models. $Z_i$ is the evidence for model $M_i$.  The evidence ratio
$Z_i/Z_j$ is referred to as the Bayes' factor $B_{ij}$ and reflects the
influence of the data.  To demonstrate the efficiency of our algorithm, we 
assume here $\pi_{ij}=1$.  
%
In this context, the models  are the waveform catalogs and the 
evidences are obtained by marginalizing over all model parameters
which are the $\{\beta_j\}$ coefficients used to construct the signal model
in equation~\ref{eq:sig} from the catalog's \pc{s}.
%
%
%
We adopt a uniform prior for $\{\beta_j\}$, in a range obtained by projecting the waveforms from
each catalog onto its corresponding \pc{s}. 
As in~\citep{Logue:2012zw}, the likelihood and corresponding evidences are computed with a nested sampling algorithm.
The model evidence is largest for the most parsimonious model that best
explains the data; $B_{ij}>1$ indicates $M_i$ is preferred over $M_j$.


\section{Identifying Binary Black Hole Merger Phenomenology}
\label{results}

\begin{figure}[t]
\sidecaption
\includegraphics{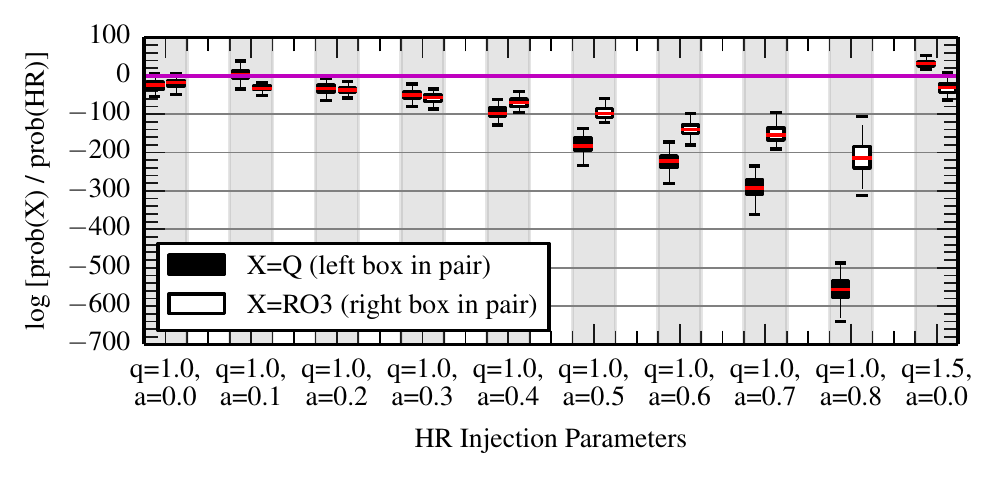}
\caption{Distribution of Bayes factors for HR waveforms. Each pair of boxes in the figure
corresponds to the sample of Bayes factors
(equation~\ref{eq:bfac_example}) for the 50 different noise realizations.
The boxes denote the interquartile range of the distribution, the red lines
indicate the median value and the whiskers show the outliers within 1.5$\times$
the interquartile range.  The $x$-axis indicates the physical parameters of the
injection performed; for the HR catalog, the mass ratio and spin magnitudes are varied. 
The two $a=0$ systems are seen to be difficult to distinguish from the Q catalog which is not
surprising since Q catalog contain waveforms for non-spinning systems.}
\label{f:HRratio}       
\end{figure}

We demonstrate the efficacy of the PCA-based Bayesian model selection with a
Monte-Carlo analysis where simulated \gw{} signals from each catalog are added to
colored, Gaussian noise, which is generated as 
in~\cite{Logue:2012zw}.  
For this proof-of-principle study we assume a single aLIGO detector operating at
design sensitivity in the ``zero-detuned, high-power''
configuration~\citep{Harry:2010zz}.  
We make the further assumptions that the
time of peak amplitude of the signal is known, that the source is optimally
oriented and located on the sky with respect to the detector and, finally, 
 that the total mass of the system is $250\,\mathrm{M}_{\odot}$.  This
choice of mass  ensures that the signals ``switch on'' below
the minimum sensitive frequency of the aLIGO noise spectrum (10\,Hz).
The physical distance of the simulated signal is scaled such that the injections
have SNR=50.
The \gw{} signals from our catalogs are injected into 50 independent noise
realizations.  
Thus, for each waveform  we obtain 50 evidence values for the waveform to belong
to one of the catalogs:  $Z_{\mathrm{Q}}$, $Z_{\mathrm{HR}}$ and
$Z_{\mathrm{RO3}}$. 

To demonstrate that  model selection  can
correctly identifying which catalog a given injection originated from,
for an injection from a given catalog $C$ we form the Bayes factors between the other catalogs
and the model $M_C$.  For example, if an injection is performed from the HR
catalog, we compute the log Bayes factors:
\begin{equation}
\label{eq:bfac_example}
\log B_{\mathrm{Q},\mathrm{HR}} =
\log Z_\mathrm{Q} - \log Z_\mathrm{HR}
\;\;\; \;\;\; \mathrm{and} \;\;\; \;\;\; 
~\log_e B_{\mathrm{RO3},\mathrm{HR}} = 
\log Z_\mathrm{RO3} - \log Z_\mathrm{HR} \,.
\end{equation}
If the algorithm correctly discriminates between the waveform
catalogs,
both $B_{\mathrm{Q},\mathrm{HR}}$ and $B_{\mathrm{RO3},\mathrm{HR}}$ will be
less than unity.

Figure~\ref{f:HRratio} summarizes the distribution of Bayes factors for HR waveforms.
The majority of the boxes lie well below zero,
indicating that the algorithm correctly identifies the HR catalog as the most
probable for these simulations, with $Z_{\mathrm{HR}} > \mathrm{max}
(Z_{\mathrm{Q}}, Z_{\mathrm{RO3}} )$.  Qualitatively similar results are
found when analyzing signals from the Q and RO3 catalogs and will be explored
more fully in a follow-up publication.



\begin{acknowledgement}
This work was supported by NSF grants PHY-0955773 and 0955825, SUPA and STFC UK.  Simulations were supported by NSF XSEDE  PHY120016 andPHY090030,  and CRA Cygnus cluster.
\end{acknowledgement}

\bibliographystyle{aps-customized}      
\bibliography{biblio}

\begin{thebibliography}{22}
\ifx \bisbn   \undefined \def \bisbn  #1{ISBN #1}\fi
\ifx \binits  \undefined \def \binits#1{#1} \fi
\ifx \bauthor  \undefined \def \bauthor#1{#1} \fi
\ifx \bjtitle  \undefined \def \bjtitle#1{\textrm{#1}}\fi
\ifx \batitle  \undefined \def \batitle#1{#1} \fi
\ifx \bctitle  \undefined \def \bctitle#1{#1} \fi
\ifx \bvolume  \undefined \def \bvolume#1{\textbf{#1}}\fi
\ifx \byear  \undefined \def \byear#1{#1} \fi
\ifx \bissue  \undefined \def \bissue#1{#1} \fi
\ifx \bfpage  \undefined \def \bfpage#1{#1} \fi
\ifx \blpage  \undefined \def \blpage #1{#1} \fi
\ifx \burl  \undefined \def \burl#1{#1} \fi
\ifx \doiurl  \undefined \def \doiurl#1{#1} \fi
\ifx \betal  \undefined \def \betal{et al.} \fi
\ifx \binstitute  \undefined \def \binstitute#1{#1} \fi
\ifx \beditor  \undefined \def \beditor#1{#1} \fi
\ifx \bpublisher  \undefined \def \bpublisher#1{#1} \fi
\ifx \bbtitle  \undefined \def \bbtitle#1{\textit{#1}} \fi
\ifx \bedition  \undefined \def \bedition#1{#1} \fi
\ifx \bseriesno  \undefined \def \bseriesno#1{#1} \fi
\ifx \blocation  \undefined \def \blocation#1{#1} \fi
\ifx \bsertitle  \undefined \def \bsertitle#1{#1} \fi
\ifx \bsnm \undefined \def \bsnm#1{#1} \fi
\ifx \bsuffix \undefined \def \bsuffix#1{#1} \fi
\ifx \bparticle \undefined \def \bparticle#1{#1} \fi
\ifx \barticle \undefined \def \barticle#1{#1} \fi
\ifx \botherref \undefined \def \botherref #1{#1} \fi
\ifx \url \undefined \def \url#1{#1} \fi
\ifx \bchapter \undefined \def \bchapter#1{#1} \fi
\ifx \bbook \undefined \def \bbook#1{#1} \fi
\ifx \bcomment \undefined \def \bcomment#1{#1} \fi
\ifx \oauthor \undefined \def \oauthor#1{#1} \fi
\ifx \citeauthoryear \undefined \def \citeauthoryear#1{#1} \fi
\ifx \texttildelow  \undefined \def \texttildelow{\symbol{126}} \fi
\def \endbibitem {}
\ifx \bconflocation  \undefined \def \bconflocation#1{#1} \fi

\bibitem[\protect\citeauthoryear{Aasi et~al.}{2013a}]{Aasi:2013jjl}
\begin{barticle}
\bauthor{\binits{J.} \bsnm{Aasi}}, \betal
\bjtitle{Phys. Rev. D}
\bvolume{88},
\bfpage{062001}
(\byear{2013}a)
\end{barticle}
\endbibitem

\bibitem[\protect\citeauthoryear{Aasi et~al.}{2013b}]{Aasi:2013wya}
\begin{botherref}
\oauthor{\binits{J.} \bsnm{Aasi}}, et al.,
2013b.
{Preprint arXiv:1304.0670}
\end{botherref}
\endbibitem

\bibitem[\protect\citeauthoryear{Abadie et~al.}{2012}]{Abadie:2012rq}
\begin{barticle}
\bauthor{\binits{J.} \bsnm{Abadie}}, \betal
\bjtitle{Phys. Rev. D}
\bvolume{85},
\bfpage{122007}
(\byear{2012})
\end{barticle}
\endbibitem

\bibitem[\protect\citeauthoryear{Acernese et~al.}{2009}]{Acernese:2009}
\begin{botherref}
\oauthor{\binits{F.} \bsnm{Acernese}}, et al.,
2009.
\url{https://tds.ego-gw.it/ql/?c=6589}
\end{botherref}
\endbibitem

\bibitem[\protect\citeauthoryear{Ajith et~al.}{2012}]{Ajith:2012az}
\begin{barticle}
\bauthor{\binits{P.} \bsnm{Ajith}}, \betal
\bjtitle{Class.Quant.Grav.}
\bvolume{29},
\bfpage{124001}
(\byear{2012})
\end{barticle}
\endbibitem

\bibitem[\protect\citeauthoryear{Andersson et~al.}{2013}]{Andersson:2013mrx}
\begin{barticle}
\bauthor{\binits{N.} \bsnm{Andersson}}, \betal
\bjtitle{Class.Quant.Grav.}
\bvolume{30},
\bfpage{193002}
(\byear{2013})
\end{barticle}
\endbibitem

\bibitem[\protect\citeauthoryear{{Clark} et~al.}{2007}]{clark:07}
\begin{barticle}
\bauthor{\binits{J.} \bsnm{{Clark}}}, \betal
\bjtitle{Phys. Rev. D}
\bvolume{76},
\bfpage{043003}
(\byear{2007})
\end{barticle}
\endbibitem

\bibitem[\protect\citeauthoryear{Harry}{2010}]{Harry:2010zz}
\begin{barticle}
\bauthor{\binits{G.M.} \bsnm{Harry}}
\bjtitle{Class. Quant. Grav.}
\bvolume{27},
\bfpage{084006}
(\byear{2010})
\end{barticle}
\endbibitem

\bibitem[\protect\citeauthoryear{{Heng}}{2009}]{heng:09}
\begin{barticle}
\bauthor{\binits{I.S.} \bsnm{{Heng}}}
\bjtitle{Class. Quant. Grav.}
\bvolume{26},
\bfpage{105005}
(\byear{2009})
\end{barticle}
\endbibitem

\bibitem[\protect\citeauthoryear{Hinder et~al.}{2014}]{Hinder:2013oqa}
\begin{barticle}
\bauthor{\binits{I.} \bsnm{Hinder}}, \betal
\bjtitle{Class.Quant.Grav.}
\bvolume{31},
\bfpage{025012}
(\byear{2014})
\end{barticle}
\endbibitem

\bibitem[\protect\citeauthoryear{Logue et~al.}{2012}]{Logue:2012zw}
\begin{barticle}
\bauthor{\binits{J.} \bsnm{Logue}}, \betal
\bjtitle{Phys. Rev. D}
\bvolume{86},
\bfpage{044023}
(\byear{2012})
\end{barticle}
\endbibitem

\bibitem[\protect\citeauthoryear{Lousto and Zlochower}{2011}]{Lousto:2010ut}
\begin{barticle}
\bauthor{\binits{C.O.} \bsnm{Lousto}},
\bauthor{\binits{Y.} \bsnm{Zlochower}}
\bjtitle{Phys.Rev.Lett.}
\bvolume{106},
\bfpage{041101}
(\byear{2011})
\end{barticle}
\endbibitem

\bibitem[\protect\citeauthoryear{Lovelace et~al.}{2012}]{Lovelace:2011nu}
\begin{barticle}
\bauthor{\binits{G.} \bsnm{Lovelace}}, \betal
\bjtitle{Class.Quant.Grav.}
\bvolume{29},
\bfpage{045003}
(\byear{2012})
\end{barticle}
\endbibitem

\bibitem[\protect\citeauthoryear{Mroue et~al.}{2013}]{Mroue:2013xna}
\begin{barticle}
\bauthor{\binits{A.H.} \bsnm{Mroue}}, \betal
\bjtitle{Phys.Rev.Lett.}
\bvolume{111},
\bfpage{241104}
(\byear{2013})
\end{barticle}
\endbibitem

\bibitem[\protect\citeauthoryear{Pekowsky et~al.}{2013}]{Pekowsky:2013ska}
\begin{barticle}
\bauthor{\binits{L.} \bsnm{Pekowsky}}, \betal
\bjtitle{Phys. Rev. D}
\bvolume{88}(\bissue{2}),
\bfpage{024040}
(\byear{2013})
\end{barticle}
\endbibitem

\bibitem[\protect\citeauthoryear{Pretorius}{2005}]{Pretorius:2005gq}
\begin{barticle}
\bauthor{\binits{F.} \bsnm{Pretorius}}
\bjtitle{Phys. Rev. Lett.}
\bvolume{95},
\bfpage{121101}
(\byear{2005})
\end{barticle}
\endbibitem

\bibitem[\protect\citeauthoryear{Reisswig and Pollney}{2011}]{Reisswig:2010di}
\begin{barticle}
\bauthor{\binits{C.} \bsnm{Reisswig}},
\bauthor{\binits{D.} \bsnm{Pollney}}
\bjtitle{Class.Quant.Grav.}
\bvolume{28},
\bfpage{195015}
(\byear{2011})
\end{barticle}
\endbibitem

\bibitem[\protect\citeauthoryear{{R{\"o}ver} et~al.}{2009}]{roever:09}
\begin{barticle}
\bauthor{\binits{C.} \bsnm{{R{\"o}ver}}}, \betal
\bjtitle{Phys. Rev. D}
\bvolume{80},
\bfpage{102004}
(\byear{2009})
\end{barticle}
\endbibitem

\bibitem[\protect\citeauthoryear{Schnetter et~al.}{2004}]{Schnetter-etal-03b}
\begin{barticle}
\bauthor{\binits{E.} \bsnm{Schnetter}}, \betal
\bjtitle{Class. Quant. Grav.}
\bvolume{21},
\bfpage{1465}--\blpage{1488}
(\byear{2004})
\end{barticle}
\endbibitem

\bibitem[\protect\citeauthoryear{Somiya}{2012}]{Somiya:2011np}
\begin{barticle}
\bauthor{\binits{K.} \bsnm{Somiya}}
\bjtitle{Class.Quant.Grav.}
\bvolume{29},
\bfpage{124007}
(\byear{2012})
\end{barticle}
\endbibitem

\bibitem[\protect\citeauthoryear{Thorne}{1987}]{Thorne:87}
\begin{bchapter}
\bauthor{\binits{K.S.} \bsnm{Thorne}},
in \bbtitle{300 Years of Gravitation},
ed. by \beditor{\binits{S.W.} \bsnm{Hawking}},
\beditor{\binits{I.} \bsnm{W.}}
(\bpublisher{Cambridge University Press},
\blocation{Cambridge, UK}, \byear{1987})
\end{bchapter}
\endbibitem

\bibitem[\protect\citeauthoryear{Vaishnav et~al.}{2007}]{Vaishnav:2007nm}
\begin{barticle}
\bauthor{\binits{B.} \bsnm{Vaishnav}}, \betal
\bjtitle{Phys. Rev. D}
\bvolume{76},
\bfpage{084020}
(\byear{2007})
\end{barticle}
\endbibitem

\end{thebibliography}

\end{document}